# Bismuth-doping Alters Structural Phase Transitions in Methylammonium Lead Tribromide Single Crystals

*Erin Jedlicka[a,†], Jian Wang[a,†], Joshua Mutch[b], Young-Kwang Jung[c], Preston Went[b], Joseph Mohammed[a], Mark Ziffer[a], Rajiv Giridharagopal[a], Aron Walsh[c,d], Jiun-Haw Chu[b], David S. Ginger[a,*]*

[a]Department of Chemistry, University of Washington, Seattle, Washington 98105, United States

[b]Department of Physics, University of Washington, Seattle, Washington 98105, United States

[c]Department of Materials and Science Engineering, Yonsei University, Seoul 03722, Korea

[d]Department of Materials, Imperial College London, SW7 2AZ, United Kingdom

AUTHOR INFORMATION

**Equal Authorship**

[†]These authors contributed equally to this work.

**Corresponding Author**

*Email: David Ginger dginger@uw.edu

**ABSTRACT**

We study the effects of bismuth doping on the crystal structure and phase transitions in single crystals of the perovskite semiconductor methylammonium lead tribromide, $MAPbBr_3$. By measuring temperature-dependent specific heat capacity ($C_p$) we find that, as Bi doping increases, the phase transition assigned to the cubic to tetragonal phase boundary decreases in temperature. Furthermore, after doping we observe one phase transition between 135 and 155 K, in contrast to two transitions observed in the undoped single crystal. These results appear strikingly similar to previously reported effects of mechanical pressure on perovskite crystal structure. Using X-ray diffraction, we show that the lattice constant decreases as Bi is incorporated into the crystal, as predicted by density functional theory (DFT). We propose that bismuth substitutional doping on the lead site is dominant, resulting in $Bi_{Pb}^{+}$ centers which induce compressive chemical strain that alters the crystalline phase transitions.

**TOC GRAPHICS**

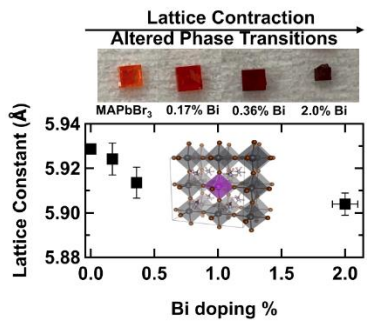

Halide perovskites have emerged as promising semiconductor materials for applications including solar cells, light-emitting diodes, photodetectors, and lasers.[1-4] They exhibit unique and tunable optoelectronic properties via facile tailoring of the chemical composition of the structure. In the archetypal perovskite $ABX_3$ crystal structure, A represents a monovalent cation species (A = $Cs^+$, $CH_3NH_3^+$ ($MA^+$), or $(NH_2)_2CH_3^+$ ($FA^+$)), B represents a divalent cation (B = $Pb^{+2}$, $Sn^{+2}$), and X represents a halide (X = $Cl^-$, $Br^-$, $I^-$). Diverse electronic and structural motifs are thereby accessible by modification of the chemical composition and the dimensionality of the material.[5] Doping provides an additional lever for changing the properties of lead halide perovskites by substituting a selected impurity into the crystal at low concentrations. Several dopant species, including $Bi^{3+}$, $Cd^{2+}$, $K^+$, $Mn^{2+}$, $Ce^{3+}$, $Yb^{3+}$, and $Eu^{3+}$,[5-13] have been studied regarding their effects on the optoelectronic properties of the lead halide perovskites. However, there have been relatively fewer studies of how doping, defects, and impurities affect the structural properties of halide perovskites crystals and nanocrystals. Doping and defects can affect temperature-dependent phase transition behavior and can alter of the lattice constant and strain.[13-15] Understanding these phenomena is important for tailoring materials properties, and for engineering materials with improved stability in varying operational conditions.

In this work, we study the effects of dopant inclusion on the crystal structure in a series of single crystals of the hybrid organic-inorganic lead halide perovskite, methylammonium lead tribromide ($MAPbBr_3$) with various amounts of bismuth doping. We chose this system for several reasons. First, the optoelectronic properties of bismuth-doped perovskites have been well characterized, among many other dopants, e.g. quenching of visible photoluminescence with corresponding near-infrared (NIR) emission, increasing conductivity, increased free carrier concentrations, and increased carrier lifetimes.[6-10,13,16] Secondly, recently developed crystallization methods for MAPbBr$_3$ report rapid growth of high purity macroscopic mm-sized single crystals with bismuth doping[6-10,17,18] which offer an ideal platform to study the intrinsic dopant effects on the crystal structure without the complication of extrinsic influences such as surfaces or grain boundaries.[19] Finally, a number of studies on the crystal structure and phase transitions of undoped MAPbBr$_3$ are available, under both atmospheric conditions and under external pressure, providing a firm literature basis for further analysis.[20-26]

Here, we investigate the effects of Bi doping on the phase transition behavior of MAPbBr$_3$ single crystals using specific heat capacity measurements ($C_p$). We show that as the Bi-doping level increases the temperature decreases for the phase-transition assigned to the cubic to tetragonal phase of MAPbBr$_3$. Additionally, upon incorporation of bismuth into the MAPbBr$_3$ crystal lattice we observe a single phase transition between 135 K and 155 K instead of the two phase transitions in that temperature range observed in the undoped MAPbBr$_3$ crystal. These changes in phase-transition behavior occur alongside a lattice contraction induced by the Bi substitutional doping at the Pb site. We compare the lattice contraction observed in X-ray diffraction (XRD) to first-principles predictions at the various doping levels for both $Bi_{Pb}^0$ and $Bi_{Pb}^+$ defect sites. We propose the incorporation of Bi occurs through a $Bi_{Pb}^+$ defect site, which induces compressive chemical strain, resulting in a lattice contraction and the changes in the phase transition behavior.

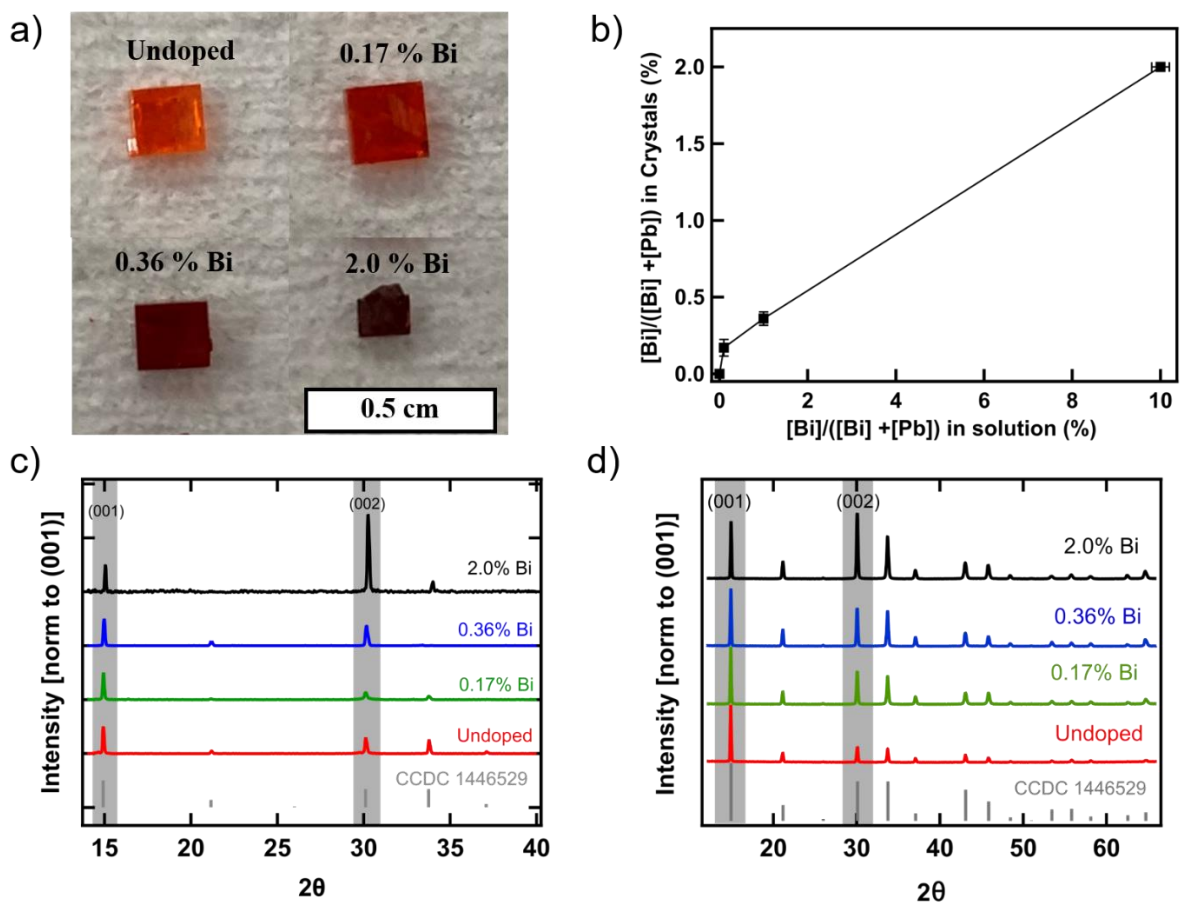

**Figure 1:** Characterization of MAPbBr$_3$ single crystals with various levels of bismuth doping. (a) MAPbBr$_3$ single crystals, (b) ratio of Bi to total amount of Pb and Bi in single crystals verses the growth solution, (c) single crystal X-ray diffraction patterns, (d) powder X-ray diffraction patterns.

We grow single-crystal MAPbBr$_3$ that incorporate various concentrations of bismuth following the inverse temperature crystallization method reported by Nayak and coworkers.[6] Detailed growth methods are provided in Supporting Information. Figure 1a shows as-prepared MAPbBr$_3$ single crystals with increasing bismuth concentration. As previously reported, we observe that bismuth doping induces strong changes in color for MAPbBr$_3$ single crystals, from translucent orange (undoped) to an increasingly darker red color (from 0.1% to 1%), then to opaque black at the highest Bi-doping (10%) levels.[6-8] This color change has been attributed to an increasing number of sub-band-gap states with increasing Bi-doping level.[6]

To quantify the bismuth-doping level in our single crystals, which can differ from the amount of Bi added to the growth solution,[6,7] we use inductively-coupled-plasma optical emission spectrometry (ICP-OES) to determine the ratio of bismuth to lead in our single crystals. Figure 1b shows a plot of the ICP-OES measured Bi concentration in the final crystal, as a function of the Bi concentration in the growth solution. As shown in Fig. 1b, we observe increasing bismuth concentration for crystals grown with higher bismuth concentrations in solution and confirm no detectable bismuth concentration in our undoped samples. We report final bismuth concentrations of 0.17%, 0.36%, and 2.0% for the 0.1%, 1%, and 10% bismuth solutions, respectively. In agreement with previous reports, we find that, at high concentrations of bismuth in the growth solution, the concentration of bismuth incorporated into the crystal is lower than the concentration of bismuth in the growth solution.[6,7]

Figures 1c and 1d show the XRD patterns, intensity normalized to the (001) plane, on both MAPbBr$_3$ single crystal and crushed powders, respectively. Both XRD show that the diffraction patterns after Bi-doping agrees with the undoped MAPbBr$_3$ reference pattern reported by Jaffe and co-workers[27] and consistent with previous reports.[6-8] We note that the (002) plane diffraction intensity increases, with respect to (001), as the bismuth level in the crystal increases. Such an increasing trend in (002) to (001) intensity in bismuth-doping level can be found in the XRD patterns in other literature reports,[6-8] though the effect was not discussed or explained. We propose that the observed increasing (002)/(001) ratio upon Bi-doping indicates increasing lattice disorder in the Bi-doped single crystals. Previous studies on undoped crystals have similarly correlated increases in the (002)/(001) intensity ratio with increasing long-range disorder as verified by far-infrared reflection (FIR) spectroscopy.[28] More ordered MAPbBr$_3$ crystals, those with less stacking faults or dislocations as quantified by a higher far-infrared reflectance intensity, also exhibit higher (001) diffraction peak intensities with respect to their (002) planes. This result is also consistent with the optical microscopy images (Fig. S1), where we clearly resolve hillock-like crystalline growing fronts in the undoped crystal. We observe fewer such growth fronts on the 0.17% Bi-doped crystal, and no such front on the 0.36% and 2.0% Bi-doped crystals suggesting less long-range order.

Next, to understand the potential impact of the Bi-doping on the phase transition behaviors, we perform temperature-dependent specific heat capacity measurements as a function of bismuth-doping concentration. Specific heat capacity measurements have been applied to study the temperature-dependent phase transition behaviors of lead halide perovskites, which have been demonstrated to depend both on the A-site alloying (FA, Cs) and X-site halide selection,[20,29] however the effects of B-site doping and substitution are less explored. Recently, Ma and co-workers showed that replacing Pb$^{2+}$ with Ni$^{2+}$ on the B-site in CsPbCl$_3$ nanocrystals altered the local structure of the doped regions and inhibited the cubic to orthorhombic phase transition.[15] Here we expect that heterovalent B-site doping, such as Bi$^{3+}$ on Pb$^{2+}$, might also alter the phase transition behavior.

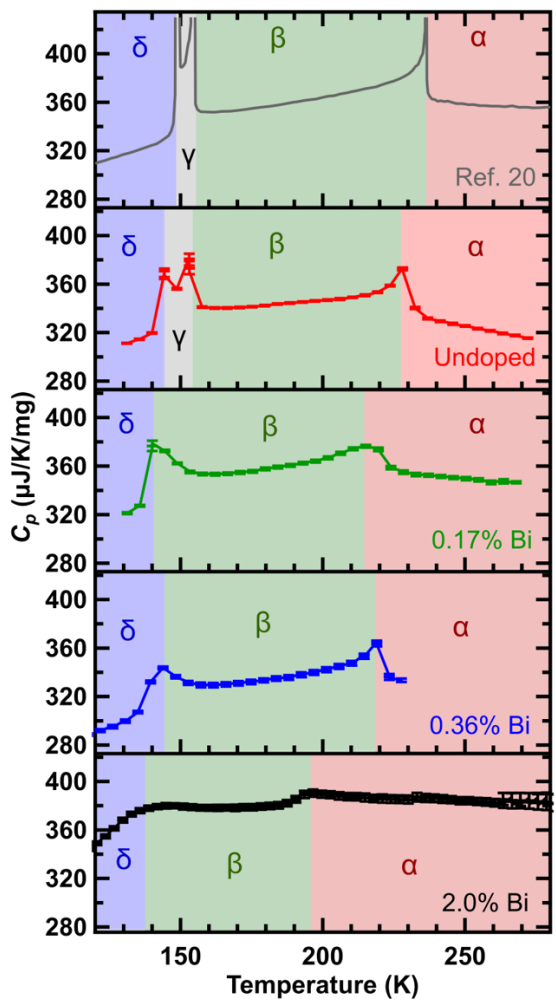

**Figure 2:** Temperature-dependent specific heat capacity ($C_p$) for various levels of bismuth doping. Phase transitions are color coded as: Red: cubic phase (α); Green: tetragonal phase I (β); Gray: tetragonal phase II (γ); Blue: orthorhombic (δ).

Figure 2 shows the heat capacity as measured using a Physical Properties Measurement System (PPMS) Dynacool™ (Quantum Design) for four MAPbBr$_3$ samples with different Bi-doping levels. Additionally, we show a reference $C_p$ vs. T plot with phase assignments as reported by Onoda-Yamamuro and co-workers.[20] In Fig. 2, the baseline represents the normal part heat capacity as a result of different vibrational modes, whereas the transitional peaks correspond to the MAPbBr$_3$ phase transition temperatures.[20] We observe three phase transitions in our undoped MAPbBr$_3$ crystals, which we determine from where the first derivative of the specific heat capacity equals zero. The peaks we measure for the undoped sample agree well with those measured by Onoda-Yamamuro et. al.

Following the previous phase transition assignments,[30,31] we identify them as: (1) the transition between α-MAPbBr$_3$, cubic (Pm3m), and β-MAPbBr$_3$, tetragonal (I4/mcm), at 228 K; (2) the transition between β-MAPbBr$_3$, tetragonal (I4/mcm), and γ-MAPbBr$_3$, tetragonal (P4/mmm), at 153 K; and (3) the transition between γ-MAPbBr$_3$, tetragonal (P4/mmm), and δ-MAPbBr$_3$, orthorhombic (Pnma), at 144 K. We assign space groups for the cubic and tetragonal phases according to studies by Poglitsch and Weber.[30] For the low temperature orthorhombic phase, we follow the assignment from Swainson and co-workers,[31] in which they suggest that Pnma space group yields more satisfactory refinement without missing symmetry, compared to previously reported Pna2$_1$ assignment. For the convenience of the reader, we have included the detailed crystal structure information for the different phases of MAPbBr$_3$ in SI Table 1. After Bi doping, we observe a few additional features: (1) the temperature of the phase transition between the cubic phase (α) to the tetragonal phase (β) shifts to lower temperatures, as seen by the expansion of the cubic phase in Figure 2, represented by the red-shaded region of the plot for the bismuth-doped samples; (2) only one low-temperature phase transition between tetragonal and orthorhombic, as seen by the disappearance of the doublet peaks near 150K likely indicating the loss of the lower temperature tetragonal phase (γ);[26] (3) at the doping level increases, phase transition peaks become less definitive but turn to gradual bumpy transition features. We note that the highest doping level (2%) crystals show greater sample-to-sample variations, as seen in Figure S2, which yet all follow the above trends qualitatively. We attribute these variations to the wider range of crystallization temperatures we observed for the 2% Bi doped single crystals. We speculate that the 2.0% Bi single crystals thus have more variation in local crystallinity which could alter the $C_p$ behavior between samples with the same nominal concentration of bismuth.

To understand the origin and implication of these observations, we compare them with similar behaviors found on the undoped MAPbBr$_3$ single crystal phase transition under external pressure. Onoda-Yamamuro et. al. have shown that (Fig. 3a, dashed lines w.r.t. top axis), for undoped MAPbBr$_3$ single crystals, as the external mechanical pressure increases, (1) the transition between the cubic phase and the tetragonal phase shifts to lower temperatures, and (2) the doublet peaks corresponding to the two tetragonal phases near 150K disappear.[21] The cause of the pressure-induced behavior is understood to be due to the unit cell volume reduction through tilting and shrinking of the PbBr$_6$ octahedra.[32] The presence of two tetragonal phases, I4/mcm (β) at higher temperature and P4/mmm (γ) at lower temperature, is unique to MAPbBr$_3$, contrasting with MAPbCl$_3$ and MAPbI$_3$ which each have only one tetragonal phase. Onoda-Yamamuro and co-workers mention that the lower temperature tetragonal (γ) phase for MAPbBr$_3$ corresponds to the P4/mmm tetragonal phase of MAPbCl$_3$ and the higher temperature tetragonal (β) phase corresponds to the I4/mcm tetragonal phase of MAPbI$_3$.[20] Khanal and co-workers attributed the formation of either P4/mmm or I4/mcm tetragonal phase to the B-X bond. In P4/mmm, the smaller octahedral volume prevents the free rotation of the MA$^+$ ions that can occur with larger octahedra in the I4/mcm space group, thus MAPbI$_3$ with a larger Pb-I bond length forms I4/mcm tetragonal phase while MAPbCl$_3$ with a shorter distance for the Pb-Cl bond length forms the P4/mmm tetragonal phase.[33] When MAPbBr$_3$ crystals are subjected to pressure, the volume of the unit cell and the Pb-Br bond length decreases.[34] This results in the disappearance of the P4/mmm (γ) phase in pressures above 43.2 MPa.[21] Chemical doping with Bi causes similar strain as the effective ionic radius of $Bi^{3+}$ (1.03Å) is smaller than $Pb^{2+}$ (1.19Å),[34] and should lead to a reduction of the lattice constant/volume, mimicking the effect of external strain, hence also inducing the disappearance of the P4/mmm (γ) phase. By closely examining the single crystal XRD data (Fig. 3b & 3c, zoom-in of Fig.1c), we observe that the diffraction plane positions shift to a higher angle, indicating a contraction of the crystal with increasing doping concentration. Notably, we only observe such a shift to higher diffraction angles on doping when performing XRD on single crystals. For the Bi-doped crystals, after crushing them into powders and performing XRD again, we observe that the diffraction planes return to the same position as the undoped crystals (Fig.S3), implying that strain builds up in the Bi-doped single crystal, and relaxes (possibly to the surface) after grinding. Previous studies also report that strain exists in Bi-doped $MAPbBr_3$ single crystals, where a broadening of the diffraction peak is observed instead of the shifting to a higher angle.[6] This result implies that a homogeneous strain exists in their system rather than the dominant compressive strain in our sample series, which might be due to subtle growth differences in the single crystals.

Fig. 3a plots the sample phase transition temperatures as a function of the lattice constants extracted from XRD. Here the top and bottom axis are registered following a previous study on undoped $MAPbBr_3$ lattice constants under external pressure.[34] The shift to a lower temperature of the cubic to tetragonal phase transition, as well as the disappearance of two tetragonal phases upon Bi doping, closely resembles the structural effects induced by an external pressure. The tetragonal to orthorhombic phase transition temperature, however, stays relatively unchanged.

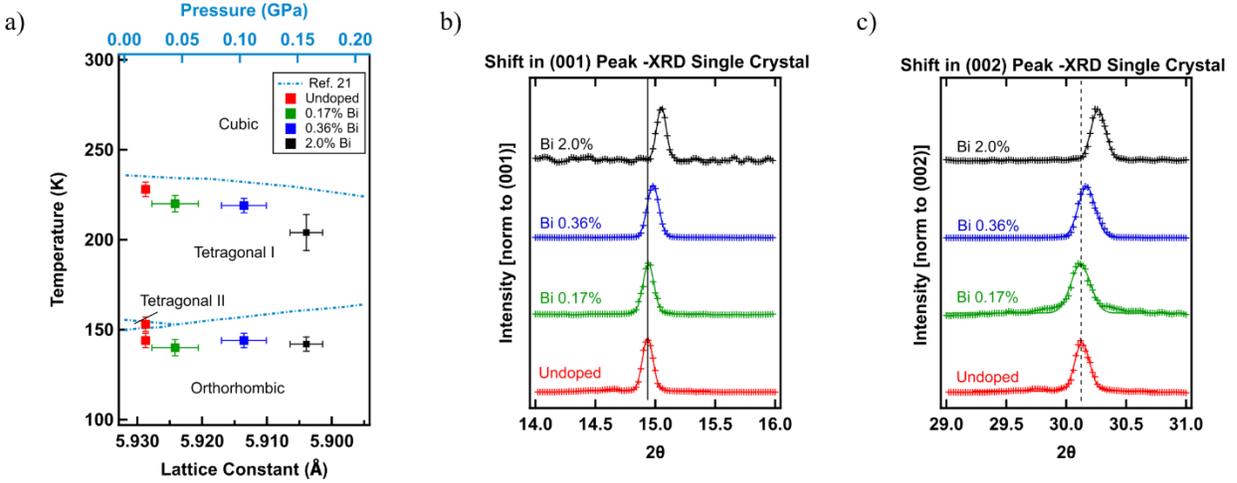

Figure 3. (a) Phase transition temperatures for pure MAPbBr$_3$ as a function of increasing pressure (blue lines, Ref. 21) and for increasing levels of Bi doping as a function of the lattice spacing. The lattice constant (bottom axis) and pressure (top axis) are registered according to a pressure-dependent MAPbBr$_3$ lattice constant study, Ref. 34 (b, c) Single crystal X-ray diffraction (XRD) showing shifts in (001) and (002) peaks to higher diffraction angles.

Finally, we perform density-functional-theory (DFT) calculations to examine the effects of the Bi dopant type on the lattice. Bi can be incorporated through a substitution on the Pb site by either Bi$^{3+}$ ($Bi_{Pb}{}^{+}$), which donates an electron into the lattice, or an effective Bi$^{2+}$ species that would result with a neutral defect species ($Bi_{Pb}{}^{0}$). Hence, we decided to consider two different charge state of Bi on Pb site. Fig. 4a shows the optimized supercell and octahedral structures. The calculated volume of the [BiBr$_6$]$^{3-}$ ($Bi_{Pb}{}^{+}$) octahedron is 10% smaller than the [PbBr$_6$]$^{4-}$ octahedron in the MAPbBr$_3$ lattice, while the [BiBr6]$^{4-}$ ($Bi_{Pb}{}^{0}$) octahedron has similar volume to the [PbBr$_6$]$^{4-}$. This indicates that shorter length of Bi-Br bonding compared to Pb-Br bonding can induce the lattice shrinkage. Since it is computationally demanding to model dilute defect concentration within DFT framework, we employed 2 × 2 × 2 supercell of cubic MAPbBr$_3$ that contains 96 atoms and adopted the thermodynamic model of defect pressure.[36] Based on the model, we calculated the change of the lattice spacing ($a_d$) as functions of $Bi_{Pb}^+$ and $Bi_{Pb}^0$ concentration following

$$a_d = a_0(1 + nv_d)^{\frac{1}{3}} \qquad (1)$$

where $a_0$ is the lattice constant of the pristine cell, $n$ is the defect concentration (defect/cm$^{-3}$), and $v_d$ is the volume of the defect (Å$^3$/defect).[36] The defect volume $v_d$ is derived according to

$$v_d = \frac{p_d * V_0}{B_0} \qquad (2)$$

and

$$p_d = -\left(\frac{\Delta(E^{defect}(V) - E^{host}(V))}{\Delta V}\right) \qquad (3)$$

where $V_0$ is the volume of the pristine cell (Å$^3$/atom), $B_0$ is the bulk modulus (18.18 GPa), and $p_d$ is the defect pressure (GPa) calculated from eq 3 using $E^{defect}$ and $E^{host}$, which are the DFT total energy of the defective and pristine cell (in eV), respectively

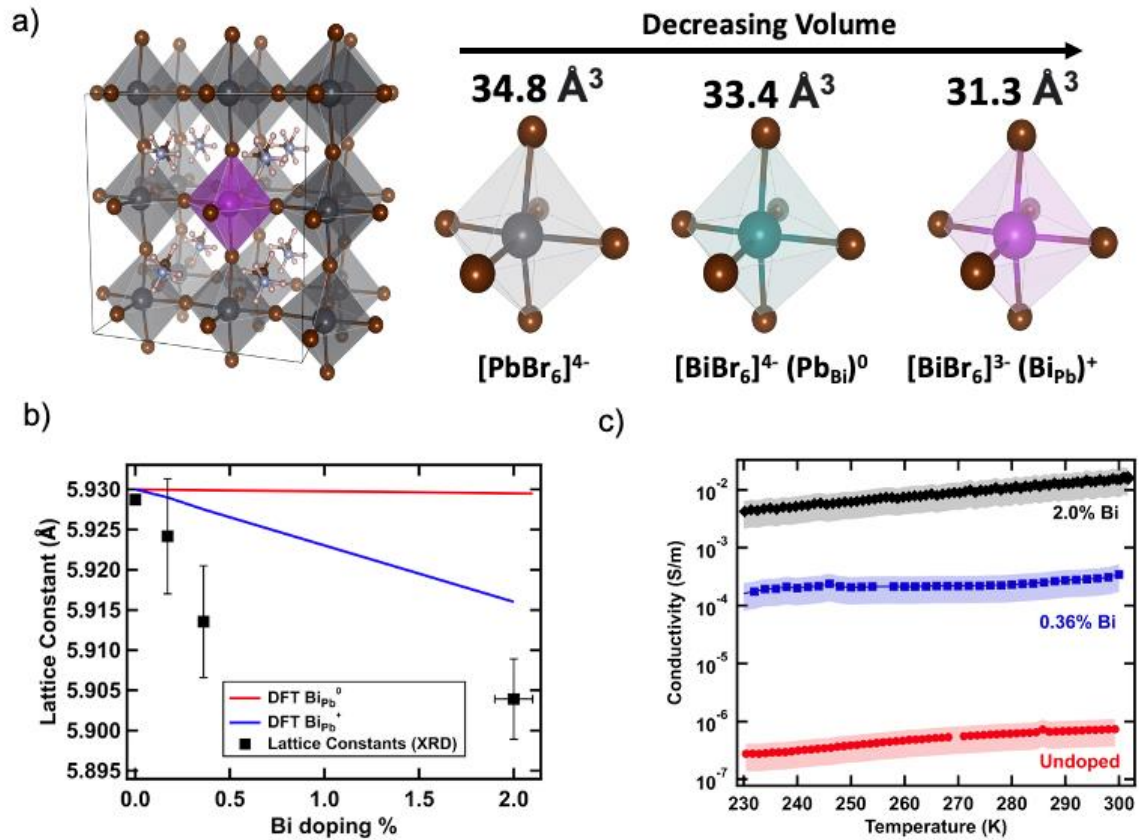

**Figure 4:** (a) Optimized structure of MAPbBr$_3$ 2 x 2 x 2 super cell geometry of octahedra in MAPbBr$_3$ lattice depending on center ions; (b) Lattice constants determined through DFT calculation (red and blue lines) and XRD (black markers); (c) Temperature-dependent conductivity measurements for single crystals.

Figure 4b shows the predicted lattice constant as a function of doping concentration. We find that the Bi$_{Pb}^+$ dopant type exhibits a negative slope, similar to our experimental results, while the Bi$_{Pb}^0$ dopant type suggests an almost invariant (slightly negative) lattice constant. Here, we note that the use of relatively small supercell with uniform Bi$_{Pb}$ distribution does not take into account gradual strain change near defect sites, which could account for some of the quantitative discrepancy with actual experiment. However, the qualitative trend of lattice compression predicted for increasing $Bi_{Pb}^+$ concentration is indeed consistent with the results from experiment. In addition, we note that $Bi_{Pb}^+$ could interact with the native negatively charged defects, of which the most probable site are MA vacancies ($V_{MA}^-$) and bromine interstitials ($Br_i^-$), forming overall neutral-charged defect pairs, i.e. $Bi_{Pb}$-$V_{MA}$ and $Bi_{Pb}$-$Br_i$.[5,9] We find that both $V_{MA}^-$ and $Br_i^-$ exhibit an increasing formation energy under pressure (Fig. S4), which implies that they have positive defect pressure that expands the lattice, i.e. they will mitigate the lattice contraction should there exist any bound pairs of $Bi_{Pb}$-$V_{MA}$ or $Bi_{Pb}$-$Br_i$.

The conductivity measurements (Fig. 4c) show that the electrical conductivity increases orders of magnitude upon Bi doping, consistent with literature reports.[8,9] This provides circumstantial evidence that $Bi_{Pb}^+$ does indeed form as a donor. From the conductivity measurements, and previous reports on carrier mobility as a function of Bi doping,[37] we estimate the electron carrier concentration to be ~$1 \times 10^{13}$ cm$^{-3}$, $2 \times 10^{16}$ cm$^{-3}$, and $1 \times 10^{18}$ cm$^{-3}$ for the undoped, 0.36% bismuth-doped, and 2.0% bismuth-doped samples, respectively. The doping efficiency, as defined by the ratio between carrier density and Bi dopant number, is therefore low (0.1~1%), which is a common feature of halide perovskites owing to efficient charge compensation mechanisms. When a charged donor is added, the system can respond by either increasing the electron carrier concentration or by forming compensating acceptor defects, for example, through methylammonium loss ($V_{MA}$), lead loss ($V_{Pb}$) or iodine gain ($I_i$). These compensating species could be distributed in the crystal or form bound complexes with Bi. The lack of quantitative agreement between the measurements and predictions for the lattice constants changes with $Bi_{Pb}^+$ in Fig 4b is likely due to the nature of these compensating species, which are unresolved at present and will be the subject of further investigation.

In summary, we have demonstrated how doping can influence the structure and phase transitions of halide perovskites. We use specific heat capacity measurements to determine the transition temperatures for various levels of Bi-doped MAPbBr$_3$ single crystals. Bi-doping alters the phase transition behavior in MAPbBr$_3$ single crystals. Comparing with the effects of external pressure on the phase transition behaviors, we note that the resemble the effects of external pressure. This observation is consistent with the experimentally observed lattice contraction upon bismuth doping. We further compare DFT calculations with experimental data and suggest a charged bismuth species replacing Pb by forming Bi$_{Pb}^+$ defects. These results provide new insight into how doping affects both the lattice structure and order in halide perovskites.

## ASSOCIATED CONTENT

**Supporting Information**. Detailed single crystal growth, characterizations (XRD, ICP-OES, Microcalorimetry, conductivity), and DFT calculations are included in Supporting Information.

## AUTHOR INFORMATION

**Notes**

The authors declare no competing financial interests.

## ACKNOWLEDGMENT

This research was supported primarily by the National Science Foundation (NSF) through the UW Molecular Engineering Materials Center, a Materials Research Science and Engineering Center (DMR-1719797). Part of this work was conducted at the Molecular Analysis Facility, a National Nanotechnology Coordinated Infrastructure site at the University of Washington which is supported in part by the National Science Foundation (grant NNCI-1542101), the University of Washington, the Molecular Engineering & Science Institute, and the Clean Energy Institute. J.W. acknowledges the funding support from the Washington Research Foundation and Mistletoe Foundation postdoc fellowships. This work was also supported by a National Research Foundation of Korea (NRF) grant funded by the Korean government (MSIT) (No. 2018R1C1B6008728). Via our membership of the UK's HEC Materials Chemistry Consortium, which is funded by EPSRC (EP/L000202), this work used the ARCHER UK National Supercomputing Service (http://www.archer.ac.uk).